\documentstyle[prl,aps,epsf,amsfonts,amssymb]{revtex}

\begin{document}
\draft
\twocolumn[\hsize\textwidth\columnwidth\hsize\csname@twocolumnfalse\endcsname
\title{Quasiparticle and Cooper Pair Tunneling in the Vortex State of
Bi$_{2}$Sr$_{2}$CaCu$_{2}$O$_{8+\delta}$}
\author{N. Morozov$^1$, L.N. Bulaevskii$^1$, M.P.
Maley$^1$, Yu.I. Latyshev$^{2\dagger}$, and
T. Yamashita$^2$}
\address{$^1$Los Alamos National Laboratory, Los Alamos, NM 87545 \\
$^2$Research Institute of Electrical Communication, Tohoku
University, 2-1-1, Katahira, Aoba-ku, Sendai 980-8577, Japan
}
\date{\today}
\maketitle
\begin{abstract}

From measurements of the $c$-axis I-V characteristics of intrinsic
Josephson junctions in Bi$_{2}$Sr$_{2}$CaCu$_{2}$O$_{8+\delta}$
(Bi-2212) mesas we obtain the field dependence (${\bf H}\parallel c$)
of the quasiparticle (QP) conductivity, $\sigma_q(H,T)$, and of the
Josephson critical current density, $J_c(H,T)$.
$\sigma_q(H)$ increases sharply with $H$ and reaches a plateau at 0.05
T$ <H< 0.3$ T. We explain such behavior by the dual effect of
supercurrents around vortices. First, they enhance the QP DOS, leading
to
an increase of $\sigma_q$ with $H$ at low $H$ and, second, they
enhance the scattering rate for specular tunneling as pancakes become
disordered along the $c$-axis at higher $H$, leading to a plateau at
moderate $H$.
\end{abstract}
]

Important information regarding pairing symmetry in cuprate
superconductors has been obtained from the quasiparticle (QP)
spectrum.  Strong evidence in favor of $d$-wave symmetry of the
superconducting order parameter has come from ARPES and STM
experiments \cite {arpes,stm}.  The effect of a magnetic field on QP
properties becomes particularly interesting because, in the nodal
regions of a d-wave gap, it probes fine details of the low-energy QP
spectrum and its alteration by supercurrents surrounding vortices.  An
increase of QP density of states (DOS) produced by the Doppler shift
of near-nodal QP energies by the vortex supercurrents is predicted to
lead to an increasing QP DOS with magnetic field \cite{gor,vol}.  This
effect was apparently observed in calorimetric measurements in HTS
\cite{moler}.  In-plane thermal conductivity measurements however have
indicated a more complex magnetic field dependence.  At low temperatures
$\kappa _{ab}(H)$ was seen to increase
sublinearly with field in agreement with predictions of the Volovik
mechanism \cite{aubin}.  At higher temperatures,
$T\! >$2--5 K.  $\kappa_{ab}(H)$ was observed in Bi-2212 to first
decrease with increasing magnetic field and then reach a plateau and
remain constant up to fields of several tesla \cite{krish}.  A similar
behavior was observed also in YBCO \cite{chiao}.  Franz \cite{franz}
attempted to explain the
high temperature plateau behavior as resulting from a compensating
effect of QP scattering from in-plane vortex disorder.  While this
provides a qualitative description of the high temperature behavior,
it fails to account for the fact that the low temperature sublinear
increase with $H$ shows no effect of a compensating scattering from
vortex disorder.  Thus, the thermal measurements have presented an
ambiguous picture of the competing contributions of vortices,
increasing DOS and scattering, to QP transport.

More direct information on QP properties can be obtained by study of
the charge transport, namely, the $c$-axis conductivity
$\sigma_{c}(H)$, in Josephson-coupled layered HTS. Recent measurements
on micron-sized mesas of Bi-2212 have demonstrated that QP
conductivity in the superconducting state can be accessed by exceeding
the Josephson critical current, switching the interlayer Josephson
junctions into the resistive state and returning to low currents on
the resistive branch \cite{lat}.  Recent theoretical development has
shown that $c$-axis QP conductivity is determined by the QP DOS and
the effective scattering rate for interlayer tunneling \cite{Vekhter}.
In this Letter we provide an unambiguous determination of the effects
of vortices on the QP DOS and on the effective scattering rate for
$c$-axis tunneling.

We study the effect of vortices on the $c$-axis QP conductivity,
$\sigma _{q}(H,T)$, by measuring the I-V characteristics in the
resistive state of small Bi-2212 mesas as a function of the magnetic
field $0<H<9$~T applied parallel to the $c$-axis.  At the same fields
we extract the dependence of the Josephson critical current density
$J_{c}(H,T)$ to obtain independent information on $c$-axis correlation
of pancake vortices.  Our main finding is that $\sigma _{q}(H,T)$
first increases rapidly with field, then reaches a plateau at
$40\;{\rm mT} <B<0.5$~T due to the dual role of vortices.  As the
field further increases, $\sigma_{q}(H)$ increases linearly, as was
found previously \cite{morozov-60T}.  We observe a moderate drop of
$J_{c}(H,T)$ with $H$ in the field range of the plateau and then a
faster drop of $J_{c}$ in fields where the linear growth of $\sigma
_{q} (H)$ takes place.  This drop indicates strong disorder in vortex
positions along the $c$-axis in high fields in agreement with
enhancement of the effective scattering rate at interlayer tunneling
responsible for plateau formation.

For our experiments we used several step-like mesa samples with
$T_{c}=73$~K and area $S=2\times 2\;\mu m^{2}$ manufactured from
high-quality Bi-2212 whiskers utilizing the 

\begin{table}
\vspace{-0.2cm}
\caption{Bi-2212 mesa samples; $S \approx 4\;\mu{\rm m}^2$,
$T_c\approx 73$~K. $R_n$ is the resistance at $T=300$~K, $N$ is the
number of intrinsic junctions, $J_c$ is the maximum critical current 
density ($H=0$), 
$\gamma$ is the
in-plane effective scattering rate, and $\sigma _q(0,0)$ is the QP
conductivity at $H=0$, $T \rightarrow 0$.}
\label{tab1}
\begin{tabular}{cccccc}
mesa & $R_n, {\rm k}\Omega$ & N & $J_{c}@{\rm 4\;K}$, ${\rm A/cm^2}$ & $
\gamma$, K & $\sigma_{q}(0,0)$, ${\rm (k}\Omega \;{\rm cm)^{-1}}$ \\
\hline
m1 & 2.1 & 42 & 830 & 32.0 & 1.5 \\
m2 & 4.5 & 90 & 500 & 31.8 & 2.1 \\
m3 & 0.85 & 17 & 1100 & 34.0 & 4.3 \\
m4 & 3.0 & 60 & 450 & --- & 1.8
\end{tabular}
\end{table}

\noindent 
Focused Ion Beam technique \cite{lat}.  In this Letter we present data obtained on 4 typical
samples (See Table I).  We performed
measurements of I-V characteristics at different magnetic fields and
temperatures in a standard He-flow cryostat, which provided
temperature stabilization better than $\pm 10$~mK. Magnetic fields up
to 9~T were applied along the $c$-axis of the mesa by a
superconducting solenoid.  The intrinsic Josephson junctions in high
quality mesas have a small shunting conductance due to QP tunneling
and thus behave as underdamped junctions with highly hysteretic I-V
dependence.  Typical multibranching I-V curves
in zero-field-cooled (ZFC) mode at $T=4$~K are shown in Fig.~1.  The
increasing current branch provides information on the critical
current, whereas the decreasing current branch corresponds to the
resistive state, where current is due to QP tunneling.  Direct
comparison of I-V curves at $H=$0~T, 0.05~T, and 0.15~T (Fig.~1)
shows that a) critical currents drop with $H$, and thus all branches
collapse to that of the resistive state, and b) variation of critical
current for different junctions increases with $H$ in comparison with
that at $H=0$.

\begin{figure}[tbp]
\vspace{-0.5cm}
\hspace{-0.5cm}
\epsfxsize = 8.5cm
\epsffile{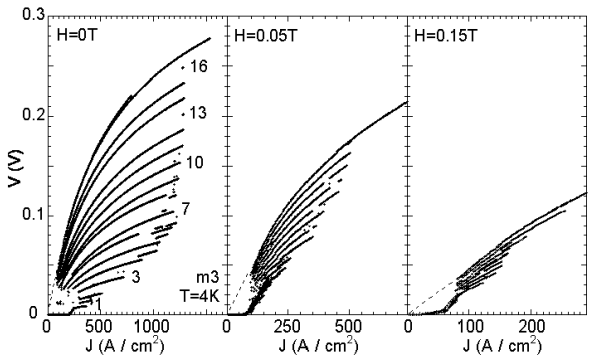}
\vspace{-0.5cm}
\caption{I-V curves for Bi-2212 mesa m3 at $H=0$,
 $H=0.05$~T, and $H=0.15$ T. For this sample
all 17 branches are resolved.
}
\label{f1}
\end{figure}

It is important that the branch of I-V curves for decreasing current, when 
all junctions are in the resistive state, is
the same for all current sweeps for both field cooling (FC), and ZFC
(Fig.~2).  This observation allows us to study the field dependence of
QP conductivity, $\sigma_q(H,T)$, measured in ZFC. The value of
average QP resistivity, $\langle \rho _{q} \rangle$, was obtained from
the all-junction resistive-state I-V curve at low currents as
described in Ref.\cite{lat}, and $\sigma _{q}=1/\langle \rho
_{q}\rangle $ was calculated.

In contrast to the behavior of $\sigma_{q}(H)$, the distribution of
critical current over junctions is different for FC and ZFC modes.
The distribution of $J_c(B,T)$ over junctions in small-area mesas
arises because a) at $T\rightarrow 0$ and $H=0$ junctions are
inequivalent with respect to $J_c$ at least due to their geometrical
position in the mesa, b) at nonzero field $J_c(B,0)$ for a junction
between layers $n$ and $n+1$ depends on vortex positions in the
layers, which vary randomly with $n$ in the case of uncorrelated
pinning \cite{fis}, and c) at nonzero $T$, jumps from the
superconducting

\begin{figure}[tbp]
\vspace{-0.5cm}
\hspace{-0.5cm}
\epsfxsize = 10cm
\epsffile{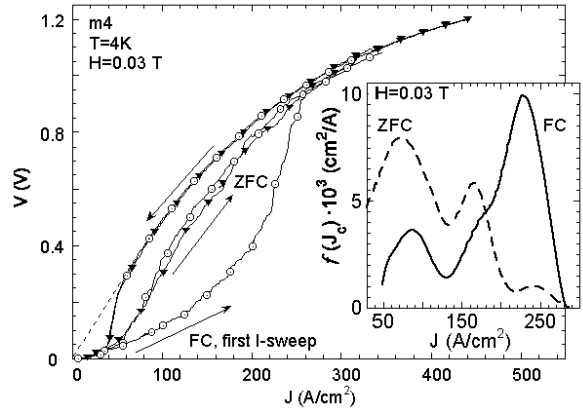}
\vspace{-2cm}
\caption{ I-V curves for  $H=0.03$~T, FC~($\circ$) and ZFC~
(\small{$\scriptstyle{\blacktriangledown}$}) modes.
In the FC mode vortices initially are better correlated, providing
larger $J_{c}$.  Switching the junctions in the resistive state
suppresses $c$-axis pancake correlations, and the system becomes
similar for both FC and ZFC. Inset: the distribution of $J_{c}$ in
junctions for ZFC (dash) and FC (line) modes. }
\label{f2}
\end{figure}

\noindent
state to a resistive one occur at different $J$ due to
thermal fluctuations \cite{Tink}.  We can find the distribution
function of the critical current density, $f(J)$, from our data as
$f(J)=d \,(V_{\uparrow}/V_{\downarrow}) / d \, J$, where $V_{\uparrow}(J)$ 
$[V_{\downarrow}(J)]$ are voltages on increasing [decreasing] currents.
The inset in Fig~2 shows smoothed distribution
functions $f(J)$ for FC and ZFC modes.  We note that average critical
current density, $\langle J_{c}(H=0.03 {\rm T})\rangle$, obtained in
the FC mode at 4 K is larger by the factor $\approx 2$ than that in
the ZFC mode.  This result is in agreement with measurements
\cite{JPR} of the Josephson plasma resonance (JPR) frequency. Below 
the irreversibility line (in the vortex glass phase) the JPR

\begin{figure}[tbp]
\epsfxsize = 8.5cm
\epsffile{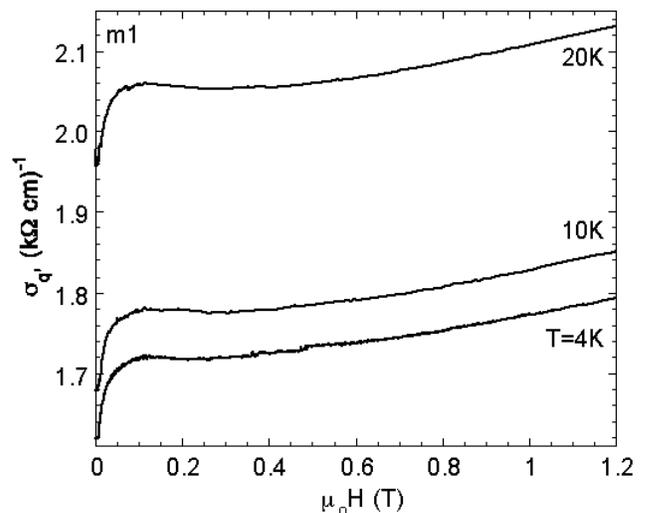}
\caption{For mesa m1 $\sigma_q(V\rightarrow 0)$ is plotted vs.
$H$ for three
different temperatures.}
\label{f3}
\end{figure}

\noindent
 frequency
in the FC mode  was found to be larger than in the ZFC mode.  A very
important new observation is that curves obtained in the
second and subsequent current sweeps in the FC mode are similar, and
they practically coincide with those observed in the ZFC mode.  This
shows that in our thin plate-like mesas suppression of Josephson
coupling by switching into the resistive state {\it irreversibly}
produces pancake disorder comparable to that produced by ramping the
magnetic field, i.e.  vortices do not change their positions when one
goes from the resistive state back into the superconducting state.

In Fig.~\ref{f3} we present our principal result showing a typical
dependence of QP conductivity as a function of magnetic field.  The
conductivity rises steeply with magnetic field, reaches a distinct
plateau, and then rises more gradually with field at higher fields.
Below we explain our experimental data using theoretical results
\cite{Vekhter} for $\sigma_q(B)$ and an estimate for $J_c(B,T)$ in
small mesas.

Let us outline the main theoretical predictions \cite{Vekhter} for QP
conductivity $\sigma _{q}(B)$ in the mixed state.  QP tunneling in a
$d$-wave layered superconductor in the presence of vortices is
determined by two competing mechanisms: $\sigma _{q}$ is proportional
to the QP DOS and inversely proportional to the effective scattering
rate for interlayer tunneling, $ \gamma _{c}$.  In-plane supercurrents
around vortices lead to a DOS increase near gap nodes proportional to
the averaged Doppler shift in the QP spectrum, $E_{H}\approx
v_{F}(B/\Phi _{0})^{1/2}$ at $\gamma \ll E_{H}\ll \Delta _{0}$, where
$\gamma $ is the impurity induced effective in-plane scattering rate
and $\Delta _{0}$ is the gap amplitude.  In the framework of the
specular tunneling model, when in-plane momentum is conserved at
tunneling, $\sigma _{q}$ will increase strongly with $B$ due to DOS
enhancement when vortices are correlated along the $c$-axis.  If $c
$-axis correlation of pancakes is absent, the effective scattering
rate $ \gamma _{c}$ increases with $B$ due to different Doppler shifts
at equivalent points in adjacent layers, suppressing tunneling.  There
is a characteristic field, $B_{\gamma }=\Phi _{0}\gamma ^{2}/\hbar
^{2}v_{F}^{2}$, estimated as $B_{\gamma }\sim 0.2-0.6$~T, which
corresponds roughly to the field at which the variation of Doppler
shift becomes comparable to the scattering rate $\gamma $.  The latter
may be estimated from the temperature dependence $\sigma
_{q}(0,T)=\sigma _{q}(0,0)(1+cT^{2})$, where $c=\pi ^{2}/18\gamma
^{2}$.  Values of $\gamma $ for mesas studied are presented in the
Table I.  It was predicted that in the quasiclassical approach, valid
at $ E_{H}\ll \Delta _{0}$, when one accounts for Doppler shifts only,
one gets $ \gamma _{c}\approx E_{H}$ in fields $B\geq B_{\gamma }$.
In this field range increase in scattering caused by Doppler shift
variation in adjacent layers compensates in $\sigma _{q}(H)$ the
increase in DOS, providing a nearly field independent $\sigma_{q}(H)$.
Corrections to this approach lead to a weak linear increase of QP
conductivity at $B_{\gamma }\ll B\ll B_0$,

\begin{equation}
\sigma_q(B)/\sigma_q(0) =C_1+B/B_0,
\end{equation}
where $B_0\approx 20$ T. Here $C_1$ depends on in-plane pancake
ordering.
The values of $C_1$ between 1.07 and 1.22 were calculated for
a 2D vortex liquid depending on the effective temperature of vortex
disorder
$T_{eff}$  \cite{Vekhter}.

\begin{figure}[tbp]
\epsfxsize = 10cm
\epsffile{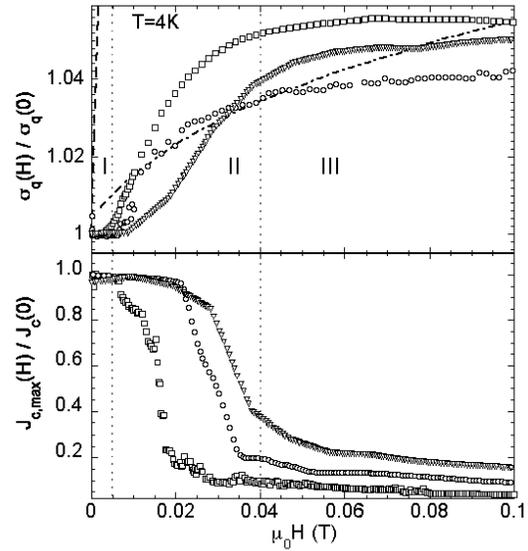}
\caption{For mesas [m1 ($\scriptstyle{\square}$), m2 ($\circ$) and m3
($\scriptstyle{\triangledown}$)] the normalized $\sigma_{q}$ (top) and
$\langle J_{c}\rangle$ (bottom) are
plotted vs. $H$. Three different regimes with respect to $H$
are marked by vertical dotted lines. The theoretical curves of
$\sigma_{q}(B) $
for correlated (dashed) and $c$-axis-uncorrelated (dash-dotted) vortex
systems
are shown for $B_{\gamma}= 0.6$ T, $\Delta_0=25$ meV, and
$T_{eff}=0.06~(\Phi_0^2s/8\pi^2\lambda_{ab}^2)$ 
\protect\cite{Vekhter}.}
\label{f4}
\end{figure}

The field dependence of the Josephson critical current in the vortex
glass state for ${\bf H}\parallel c$ comes from the dependence of
$J_{c}(B,T)=J_{c}(0,T) \langle \cos \varphi _{n,n+1}({\bf r})\rangle $
on the gauge-invariant phase difference $\varphi _{n,n+1}({\bf r})$
induced by pancakes when they are misaligned in the layers $n$ and
$n+1$ due to random pinning.  Correlations of pancakes in neighboring
layers, leading to lower energy, are induced by magnetic coupling of
pancakes and by Josephson interlayer coupling; both tend to align
vortices along the $c$-axis.  Maximum $J_c$ in the vortex glass phase
may be achieved by adjustment of pancake positions to minimize both
Josephson and pinning energy \cite{kosh}.  This effect is negligible
for the ZFC mode and, as shown above, for the FC mode after switching
from the resistive state where Josephson coupling is absent.  For the
critical current $J_c(B,0)$ in mesas with the size $L\ll
\lambda_J^2/a$, the relations

\begin{eqnarray}
&&\frac{\langle J_c^2(B,0)\rangle}{J_c^2(0,0)}=
\int \frac{d{\bf r}d{\bf r'}}{L^4}
\langle\exp[i\varphi_{n,n+1}({\bf r})-i\varphi_{n,n+1}({\bf
r'})]\rangle,
\label{ph} \\
&&\varphi_{n,n+1}({\bf r})=\sum_i \phi_v({\bf r}-{\bf r}_{ni})-
\phi_v({\bf r}-{\bf r}_{n+1,i})
\nonumber
\end{eqnarray}

\noindent hold \cite{fis}.  Here $a=(\Phi_0/B)^{1/2}$ is the intervortex
distance, $\lambda _{J}$ is the Josephson length, $\sim 1\;\mu m$ in
typical mesas, $\phi_v({\bf r})$ is the polar angle of the point ${\bf
r}$, and ${\bf r}_{ni}$ is the coordinate of pancake $i$ in the layer
$n$.  In Eq.~(\ref{ph}) the integrand,
$\langle\exp[i\varphi_{n,n+1}({\bf r})-i\varphi_{n,n+1}({\bf
r'})]\rangle$, drops with $|{\bf r}-{\bf r'}|$ on the scale $a$ at
$a\ll L$.  Then we estimate $\langle J_c(B,0)\rangle<\langle
J_c^2(B,0)\rangle^{1/2}\approx J_c(0,0)(\Phi_0/BL^2)^{1/2}$.  Thermal
fluctuations cause jumps into the resistive state at $J<J_c(B,0)$ and
diminish additionally the average critical current.  When the
Josephson energy, $\Phi_0\langle J_c(B,0)\rangle L^2/2\pi c$, becomes
less than $\approx 10T$ (see Eq.~(6.18) in Ref.~\cite{Tink}), thermal
fluctuations practically eliminate the superconducting part of I-V
curves.  For our mesas this occurs in fields $\gtrsim$ 0.01 T at
$T=4$~K.

Now we explain our experimental results using these theoretical
results.  In Fig.~\ref{f4} we plot the low-field part of the $\sigma
_{q}(H)$ dependence together with extracted $J_{c,max}(H)$
for three different samples.  Shown here are also theoretical curves
calculated for $\sigma_{q}(B)$ in the case of $c$-axis-correlated and
uncorrelated vortex states \cite{Vekhter}, indicating clearly that
$c$-axis disorder is responsible for strong suppression of
$\sigma_{q}$ at high fields.  The experimentally observed
$\sigma_{q}(H)$ can be divided into three segments as magnetic field
increases as shown in Fig.~4.  These segments correspond to
appropriate parts of the field dependence of the critical current
density, $ J_{c,max}$, obtained in the ZFC mode, presented
in the lower part of Fig.~4.  At very low fields, $H\lesssim 5$~mT,
(Segment I) there is practically no change in $\sigma _{q}$ and in
$J_{c,max}$.  This regime corresponds to the Meissner state
of the sample.  Note that, despite the suppression of supercurrent
along the $c$-axis by applied transport current in the resistive
state, the in-plane Meissner currents persist, preventing pancake
vortex entrance into the sample.  Only when the magnetic field $H$
exceeds $\sim 5$~mT do pancakes enter into the sample.  In the
intermediate field range, $5\leq H<40$~mT, (segment II) the
experimental results for $\sigma_{q}(H)$ lay slightly above the curve
for uncorrelated vortices but well below the curve for the
$c$-axis-correlated vortex state.  This indicates that weak $c$-axis
vortex correlations are present in this field range.  Weakly
correlated pancakes affect $J_{c}$ much more strongly than
$\sigma_{q}$ , because increase in $B$ leads to decrease of $J_{c}$ (
the fluctuations of the phase difference increase and $\langle \cos
\varphi _{n,n+1}( {\bf r})\rangle $ drops), while for $\sigma_{q}$ the
increase of DOS is compensated partly by an increase of $\gamma_{c}$.
As the field continues to increase above 40~mT (segment III) increase
of $\sigma_{q}$ with $H$ is slowing down due to a stronger decrease of
$c$-axis correlations and, respectively, stronger increase of
$\gamma_{c}$.  This leads to formation of the plateau (or even dip) in
the range $40<H<300$~mT.

This scenario is confirmed by the behavior of $J_{c,max}(H)$.  A sharp
drop in $c$-axis correlations above 40~mT was observed in Bi-2212
single crystals as the second peak effect.  It was shown by neutron
scattering \cite{forgan} and JPR frequency study \cite{marat} that
here the vortex system decomposes into a $c $-axis-uncorrelated vortex
glass.  Accordingly, the $c$-axis critical current density, $J_{c,max}$,
drops rapidly, as seen clearly in the insert in
Fig.~\ref{f1}.  At fields above 0.5~T the dependence $\sigma_{q}(H)$
changes again, switching to the linear growth, observed and discussed
in our previous work \cite{morozov-60T}.  We obtain $B_{0}\sim 20$~T
and $C_{1}\sim 1.04$.  We see that $C_{1}$ is smaller than that
calculated for the vortex liquid, and this means that in-plane
ordering in the vortex glass is better than in the liquid.

To conclude, we observed a plateau in $\sigma_q(H)$, which formed as a
result of the compensation of the increase of QP DOS due to Volovik
effect by the increase of the effective scattering rate for tunneling
due to $c$-axis pancake disorder.  The field dependence of the QP
interlayer conductivity is sensitive to the structure of vortex state.
We found that in the vortex glass adjustment of vortex positions to
the Josephson interlayer coupling depends on the way this vortex state
was achieved.  Critical current in the FC mode is $\sim 2$ times
larger than that obtained in the ZFC mode or in the FC mode after
switching from the resistive state.  The $d$-wave pairing model in the
clean limit with resonant intralayer scattering and significant
contribution of specular interlayer tunneling provides an explanation
for the $c$-axis electrical transport in the vortex state of Bi-2212
crystals at low temperatures and low magnetic fields.

We thank A.M. Nikitina for providing us with Bi-2212 single crystal
whiskers,  Ch.~Helm, A.E.~Koshelev, and I.B.~Vekhter for useful
discussions, and J.Y.~Coulter for the  technical assistance.
This work was supported by the Los Alamos National
Laboratory under the auspices of the U.S. Department of Energy, CREST,
the Japan Science and Technology Corporation, and the Russian State
Program on HTS under grant No.~99016.


\begin{references}
\vspace{-1cm}
\bibitem[\dagger]{address}  Permanent address: Institute of Radio
Engineering
and Electronics RAS, 11 Mokhovaya, 103907, Moscow.

\bibitem{arpes}  H.~Ding {\it et al.}, Phys. Rev. Lett. {\bf 74}, 2784
(1995); J.~Mesot {\it et al.}, {\it ibid.} {\bf 83}, 840 (1999).

\bibitem{stm}  Y. DeWilde {\it et al.}, Phys. Rev. Lett. {\bf 80}, 153
(1998).

\bibitem{gor}  L.P. Gor'kov and P.A. Kalugin, JETP Lett. {\bf 41}, 253
(1985).

\bibitem{vol}  G.E. Volovik, JETP Lett. {\bf 58}, 6 (1993).

\bibitem{moler}  K.A. Moler {\it et al.}, Phys. Rev. Lett, {\bf 73},
2744 (1994); J.W. Loram {\it et al.}, J. Phys. Chem. Sol. {\bf 59}, 2091
(1998).

\bibitem{krish}  K. Krishana {\it et al.}, Science {\bf 277}, 83 (1997);
H.~Aubin {\it et al.}, {\it ibid.} {\bf 280}, 9a (1998); Y. Ando {\it et
al.}, cond-mat/9812265, 1999.

\bibitem{chiao}  M. Chiao {\it et al.}, Phys. Rev. Lett. {\bf 82}, 2943
(1999).

\bibitem{aubin}  H. Aubin {\it et al.}, Phys. Rev. Lett. {\bf 82}, 642
(1999).

\bibitem{franz}  M. Franz, Phys. Rev. Lett. {\bf 82}, 1760 (1999).

\bibitem{lat}  Yu.I. Latyshev {\it et al.}, Phys. Rev. Lett. {\bf 82},
5345 (1999).

\bibitem{Vekhter}  I.~Vekhter {\it et al.}, Phys. Rev. Lett. {\bf 84},
1296 (2000).

\bibitem{morozov-60T}  N.~Morozov {\it et al.}, Phys. Rev. Lett. {\bf
84}, 1784 (2000).

\bibitem{fis}M.V. Fistul, JETP Lett. {\bf 52}, 192 (1990).

\bibitem{Tink}M.~Tinkham, {\it Introduction to Superconductivity},
(McGraw-Hill, 1996).

\bibitem{JPR}  Yu. Matsuda, {\it et al.}, Phys. Rev. Lett. {\bf 78},
1972 (1997).

\bibitem{kosh}  A.E. Koshelev {\it et al.}, Phys. Rev. B {\bf 53}, 2786
(1996).

\bibitem{forgan}  R. Cubitt and E.M. Forgan, Nature {\bf 365}, 407
(1993).

\bibitem{marat}  M.B.~Gaifullin {\it et al.}, Phys. Rev. Lett. {\bf
84}, 2945 (2000).
\end{references}
\end{document}